\documentclass[DIV12,11pt]{scrartcl}
\usepackage[utf8]{inputenc}
\usepackage[english]{babel}
\usepackage{pdflscape}
\usepackage{tikz}
\usepackage[absolute,overlay]{textpos}
\usepackage{authblk}

\usepackage[margin=1in]{geometry}

\usepackage{amsmath,blkarray}

\usepackage{amsthm,amssymb,amsmath}
\usepackage{lipsum}
\usepackage{multirow}
\usepackage{multicol}
\usepackage{graphicx}
\usepackage{algorithmic}
\usepackage{comment}
\usepackage{caption}
\usepackage{color}
\usepackage{setspace}
\usepackage{array}
\newcolumntype{L}[1]{>{\raggedright\let\newline\\\arraybackslash\hspace{0pt}}m{#1}}
\newcolumntype{C}[1]{>{\centering\let\newline\\\arraybackslash\hspace{0pt}}m{#1}}
\newcolumntype{R}[1]{>{\raggedleft\let\newline\\\arraybackslash\hspace{0pt}}m{#1}}
\usepackage{tabulary}

\setkomafont{title}{\normalfont \LARGE \bfseries}
\setkomafont{section}{\normalfont \Large \bfseries \boldmath}
\setkomafont{subsection}{\normalfont \large \bfseries}
\setkomafont{subsubsection}{\normalfont \normalsize \bfseries}
\setkomafont{descriptionlabel}{\itshape}
\setkomafont{paragraph}{\normalfont \bfseries \boldmath}

\newtheorem{theorem}{Theorem}
\newtheorem{lemma}[theorem]{Lemma}

\usepackage[
backend=biber,
style=apa,
]{biblatex}

\usepackage{fancyhdr}
\fancypagestyle{myplain}
{
  \fancyhf{}

  \fancyfoot[C]{\thepage}
}

\newcommand\blfootnote[1]{%
  \begingroup
  \renewcommand\thefootnote{}\footnote{#1}%
  \addtocounter{footnote}{-1}%
  \endgroup
}

\author[1]{Mustafa Kaan Topaloğlu}
\author[1]{Banu Kabakulak*}
\affil[1]{Department of Industrial Engineering, İstanbul Bilgi University, \.{I}stanbul, Turkey}

\title{A Two-Phase Method for Production Planning and Machine Speed Optimization Problem }
\date{ }

\begin{document}

\maketitle \blfootnote{*Corresponding author: Banu Kabakulak, e-mail: banu.kabakulak@boun.edu.tr}

\thispagestyle{empty}
\begin{abstract}

\vspace{-20mm}

Textile industry is becoming a highly competitive area with the increase in demand for textile products. Since expanding the production capacity is not always feasible, optimizing the existing system is more practical. In particular, we consider a felt production system of a textile factory operating in Turkey in this study. We aim to minimize the production costs by optimizing machine operating speeds as well as building an efficient production lot sizing plan within the planning horizon. In this direction, we propose the Lot Sizing and Machine Speed (LSMS) nonlinear model to determine the optimal unit processing times and production quantities while minimizing the work-in-process and end item inventories by changing the machine operating speeds dynamically according to the demands. Since LSMS nonlinear optimization problem is NP-hard, we design a $\emph{Two-Phase heuristic}$ which iteratively processes a linear programming model by utilizing a commercial solver at each phase. We intensively test our $\emph{Two-Phase heuristic}$ via randomly generated demand, planning horizon and machine-hour capacity scenarios. Our computational experiments show that the introduced $\emph{Two-Phase heuristic}$ can find the local-optimal results in acceptable amount of time. 

\textbf{Keywords:} Nonlinear mathematical modelling, $\emph{Two-Phase}$ heuristic, lot sizing, machine speed optimization

\end{abstract}

\newpage

\section{Introduction and Literature Review}

Industries are trying to adapt themselves to the increase in demand in their products, which is caused by the increase in the population worldwide. Today, the textile industry, which is used in automotive, ready-made clothing, packaging, construction, and many similar areas, has become a competitive industry due to the increasing demands in the areas where it is used. The fact that most textile factories are formerly established and do not have major production systems, the competition possibilities of these factories are limited. 

The felt production system examined in this study is optimized by minimizing the total operating cost, additionally making a production plan via optimizing unit processing times and the production amounts of the machines. At the end of the optimization, it is envisioned that competition level of the production system will increase while keeping efficiency level of the production system high, as well as increasing the total production capacity of the system. 

Production plan plays a crucial role in every factory and in the production system. Success and profitability of the production system depends on a good plan. There have been many researches in developing and generalizing the idea of the production plan with many variables. To achieve the least possible cost in a given planning horizon whilst satisfying all the requirements of the system, \cite{GMH1955} examine a production schedule of a system and shape up a production plan by constructing different scenarios with requirements. Modigliani and Hohn form production intervals throughout the planning horizon and show that the optimal production plan consists of series of optimal production intervals. In later years, \cite{GMHalg1979} generalize on Modigliani and Hohn's work and provide a computational algorithm called Generalized Modigliani and Hohn (GMH) Algorithm. \cite{FLORIAN1971} hypothesized the idea of deterministic production plan with non-convex cost and capacity constraints, with finite number of periods and without varying capacity from period to period. While providing two scenarios where backlogging allowed and not allowed, \cite{FLORIAN1971} use dynamic programming to generate the shortest path route to find optimal solutions.

To design good production parameters, such as lead time and lot size, in a hierarchical production plan environment, \cite{GANSTERER2014206} propose a frame-work which is optimized by simulation-based optimization methods. \cite{MARTINEZ2016554} construct a linearized model for molded pulp packing industries production plan with the objective of minimizing total set up cost and inventory costs, to achieve this planning horizon is splitted into periods, and periods divided into slots. In their research, \cite{naeem2013} develop a dynamic programming model for minimizing the cost of manufacturing and remanufacturing system while minimizing inventory of the system and the backlog. The system in \cite{naeem2013} has finite number of stages within the planning horizon. With the proposed deterministic and stochastic dynamic programming model, \cite{naeem2013} produce policies how much to produce at each stage. In their work, \cite{naeem2013} show that the optimal production plan achieved by developing optimal policies for stages throughout the planning horizon. In work of \cite{Maria2003}, goal programming model is developed and for the multi-objective production plan problem, a Linear Psychical Programming is proposed. \cite{Yu2011} generate two hybrid evolutionary methods named as gradient-based SPEA2 and gradient-based NSGA\_II, to optimize a nonlinear multi-objective production plan in mineral processing. In creation of production plan for continuous steel casting, \cite{zhu2010intelligent} propose a hybrid optimization model which minimizes the queues, starting time of the operations and conflicts of tasks by combining genetic algorithm and parallel-backward inferring algorithm, that works consecutively solve the model. To evaluate the model, \cite{zhu2010intelligent} create a simulation model based on cellular automata (CA). \cite{LI2009286} propose a hybrid cell evaluated genetic algorithm (CEGA) which merges genetic algorithm with fractional factorial design for creating a simulation based optimization method for a remanufacturing system's production plan in a stochastic environment with the objective of maximizing profit. With the proposed method, CEGA is more effective method to search global optimum than other simulation methods. In a very sensitive production system where the raw materials holding cost is high and raw materials have different processes to make different products, such as sawmill system, \cite{Thomas2002} propose multi period production plan model (M3PM) with the objective of maximizing the total net revenue. Proposed M3PM method consists of two segments: first one is Linear Programming (LP) algorithm named as resource coordinator which creates optimal levels of activities that meets the constraints, second one is a simulation model which constructs columns (activities) by using Dynamic Programming (DP) for the LP model. 
 
Not all production systems are discrete. \cite{PERES2000} propose methods for the continuous-time production systems, where demands are satisfied continuously as well as raw materials arrives to the system successively. To generate a solving procedure for this problem, \cite{PERES2000} discuss that planning horizon should be breakdown into a periods and production rates of the machines should be allowed to change to reduce the cost within the period, called at \emph{switching times}, not at the end of the period. The method, named as two-step iterative procedure, consists of two stages, at first stage production rates are optimized, at second stage switching times within the periods are optimized. Objective of this method is that minimizing the generated cost from backorder and inventory keeping within the production system. Main difference between our study and the problem in \cite{PERES2000} is that, our system is a discrete system with allowed work-in-process (WIP) accumulation, not a continuous one nor WIP disallowed. Therefore there is no need of changing the machine speed within a period. Additionally, our mathematical model is a nonlinear programming (NLP) formulation due to the machine speed constraints, whereas \cite{PERES2000} have a nonlinearity because of objective coefficients. 

Production planning contains a lot-sizing problem (LSP), that is deciding how many products to produce in each period. Although in our production system there is a cutting operation, our study does not include the cutting stock problem (CSP) since the aim of our study to minimize cost while changing the unit processing times, thus we only consider the LSP within our production plan. In  \cite{GRAMANI2009219}, both LSP and CSP problems are considered for production planning in a furniture industry. \cite{GRAMANI2009219} say that, treating LSP and CSP problems separately, increases the global cost, hence propose a Lagrangian-based heuristic to combine and optimize the lot-cut problem (LCP) while generating lower cost. 

One form of production planning is aggregated production plan (APP) which includes laying off, hiring and overtime. By combining dynamic programming and metaheuristic methods such as bat and bee algorithms, \cite{luangpaiboon2017} proposes an a method named as \emph{Two-Phase} Approximation Method. The method first uses bat and bee algorithms to find feasible solutions for the APP problem, then this dynamic programming method choose the optimal algorithm for the problem. For finding an optimal APP, \cite{stephen2004} come up with a stochastic robust optimization model with a nonlinear objective. To solve the problem as a linear problem, \cite{stephen2004} introduce two deviational decision variables to the objective function. Offered robust optimization method is used later on forming of multi-site production plan for a company within the unclear environment with the objective of minimizing total cost. \cite{WANG20143069} find out that particle swarm optimization (PSO) is insufficient method for optimizing an APP with mixed integer linear programming model and propose a modified PSO (MPSO) which has sub-particles in contrast to the standard PSO. In measuring the efficiency of MPSO, \cite{WANG20143069} use Genetic Algorithm (GA) and standard PSO and show that modification provides better results in accuracy, converge time and reliability. In creating an APP in a continuous-time environment \cite{gganesh2005} introduce a hybrid method for finding a global optimum which combines GA and simulated-annealing (SA) named as GA-SA. \cite{gganesh2005} optimize the continuous-time APP with the objective of minimizing the total cost while considering randomly generated production rates and planning horizons within the APP. In their work, \cite{gganesh2005} use GA, SA and hybrid GA-SA separately and conclude that proposed hybrid method works best among all three methods. \cite{CHAKRABORTTY2015366} use a modified and specific variant of PSO named as linear reduction of inertia weight to solve the multi-period, multi-product integer linear programming APP problem with possibilistic environment (PE) and the objective of minimizing total cost. To solve the uncertainty which comes from PE, \cite{CHAKRABORTTY2015366} propose fuzzy triangular membership function. With a case study, proposed PE-PSO method showed better qualities than GA and fuzzy based GA.

\cite{MODARRES20161074} set up a linear multi-period, multi-objective APP for energy saving with some uncertain parameters and objectives of minimizing energy consumption, operating cost and carbon emission while considering production, inventory and backlog. Proposed solution method first uses goal attainment technique and then utilizes robust optimization method with aggregate constraint approach. Another approach for forming a multi-objective multi-period APP with uncertainty according to seasons, mixed integer nonlinear programming (MINLP) model of \cite{GOLI2019} with objectives of maximizing the customer satisfaction and minimizing the total cost. \cite{GOLI2019} create four seasons with different uncertain demands and design the model such that it can optimize every season separately. To deal with the uncertainty, robust optimization is introduced to the model. After clarifying the uncertainty, goal programming with the algorithms multi-objective invasive weed optimization algorithm and non-dominated sorting genetic algorithm are introduced to solve the model. After the optimization with applying Taguchi Method, the efficiency is improved by finding the best set of parameters of the proposed model.

Key processes like mining, drilling or, in general extracting raw materials, happens outside of the factory environment. Hence, not all production plans have to be in a factory environment. In organizing a production plan in an open-pit mine which consists of set of blocks with different attributes, \cite{SHISHVAN2015825} discuss a new metaheuristics which derives from Ant Colony Optimization to maximizing the total profit. \cite{SHISHVAN2015825} show that the proposed metaheuristics are Max-Min Ant System and the Ant Colony System  can solve the open-pit mine problem with any type of nonlinear constraints and objective function. Establishing a production plan for petroleum refineries, \cite{Sérgio2005} propose a extended model that is a multiperiod MINLP problem with uncertainty which is derived from the Pinto and Moro's model. The objective function is maximizing the profit while taking inventory levels, operation and raw materials costs into consideration and solved by using MINLP algorithms. In a uncertain environment with significant changes in market demands, companies may use reconfigurable manufacturing system (RMS) to adapt to changes. To utilize RMS effectively \cite{abbasi2011} proposes a MINLP model with stochastic order environment.The RMS model optimizes the production tasks within the planning horizon as well as optimizing the batch sizes to minimize the inventory. The RMS model objective is maximizing profit while considering production costs, inventory costs and changeover costs from single-task to multiple-task production. Proposed solving procedure is finds optimal/near-optimal output using GA. \cite{ZHAO20161} introduce a MINLP model with the objective of maximizing the overall profit for ethylene plant while considering inventory, processing operations, materials, energy consumption and energy utilization which is more efficient than the base model. \cite{ZHAO20161} create two scenarios with same planning horizons and different number of periods and noted that as the number of periods increases (as the system becomes continuous time), the quality of the solution and profit increases, the size of the model increases, thus increases the computational time. 

In this study, for developing and optimizing the felt production system, we proposed a NLP production model with the objective of minimizing the total production cost which consists of processing cost, inventory holding cost, transportation cost of the inventories and energy consumption of the machines. We aim to minimize the objective function with lot sizing which will be achieved by rearranging the machine speeds (unit processing times) at each period according to the demand. To solve the proposed mathematical model in polynomial time we developed a heuristic named as \emph{Two-Phase} method to get a near-optimal solution for the proposed mathematical model. We tested the system under three different scenarios generated with three key parameters which are planning horizon of the production, maximum inventory levels and plant capacity.


The contributions of this research to the literature can be summarized as follows:
\begin{itemize}
\item We propose a NLP model for the felt production system which creates a production plan by changing machine speeds at the beginning of each period within the planning horizon.
\item We develop a linear programming based \emph{Two-Phase} heuristic, which works at polynomial time to get a near-optimal solution for the nonlinear model.
\item We test the performance of \emph{Two-Phase} heuristic on randomly generated instances. The results indicate that \emph{Two-Phase} can find near optimal production plan and machine speeds for real size instances in acceptable time.
\end{itemize}

\section{Problem Definition}

In this study, we focus on the machine speed optimization and lot sizing problem in an existing felt production facility. The proposed production plan aims to minimize the total cost by decreasing the inventory levels, deciding what to produce and when to produce by dynamically changing the machine speeds (unit processing times of the machines) within the planning horizon $T$. Planning horizon $T$ is divided into periods $t$ to create a efficient plan for deciding the unit processing times of the machines as well as satisfying demands of the considered periods. 

In the felt production system, there are three different machines which are named as Production Line 1 (PL1, $m = 1$), Production Line 2 (PL2, $m = 2$) and Cutting Machine (CM, $m = 3$). In PL1, there are three operations, these are mixing, carding and needling. These mechanical operations turn the raw material into the compact felt products. At the end of the PL1, the felt is rolled into a cylinder. In PL2, compact felt products go through chemical process by chemicalization operation and free conveyor dryer machine. Similar to PL1 chemical felts are rolled into a cylinder at the end of the PL2. CM is cutting the felt product into the plaque shapes. According to these machines, there are four different types of products with respect to their production line and machine routing. These are: Non-Chemical Cylinder, Non-Chemical Plaque, Chemical Cylinder and Chemical Plaque. 
\begin{itemize}
    \item \emph{Non-Chemical Cylinder ($i = 1$): }Raw material goes through only PL1 and becomes an end item.
    \item \emph{Non-Chemical Plaque ($i = 2$): }Raw material goes through  PL1 and CM respectively and becomes an end item.
    \item \emph{Chemical Cylinder ($i = 3$): }Raw material goes through PL1 and PL2 respectively and becomes an end item.
    \item \emph{Chemical Plaque ($i = 4$): }Raw material goes through  PL1, PL2 and CM respectively and becomes an end item.
\end{itemize}

Raw material processing in the system always starts in PL1. Production can not start in PL2 nor in CM, since the inputs to these machines inputs are basic compact felts which are only produced in PL1. Machine routing sequence defines the product types, thus it is not possible to change the production routing. For example it is not possible to produce a Chemical Plaque with a sequence of PL1-CM-PL2, since PL2 takes basic compact felts as a cylinder unit, not as a plaque unit. This makes the production system a flow-shop. In Figure 1, the layout of the production system can be seen.

\begin{figure}[!h]
\begin{center}
\includegraphics[width=1\columnwidth]{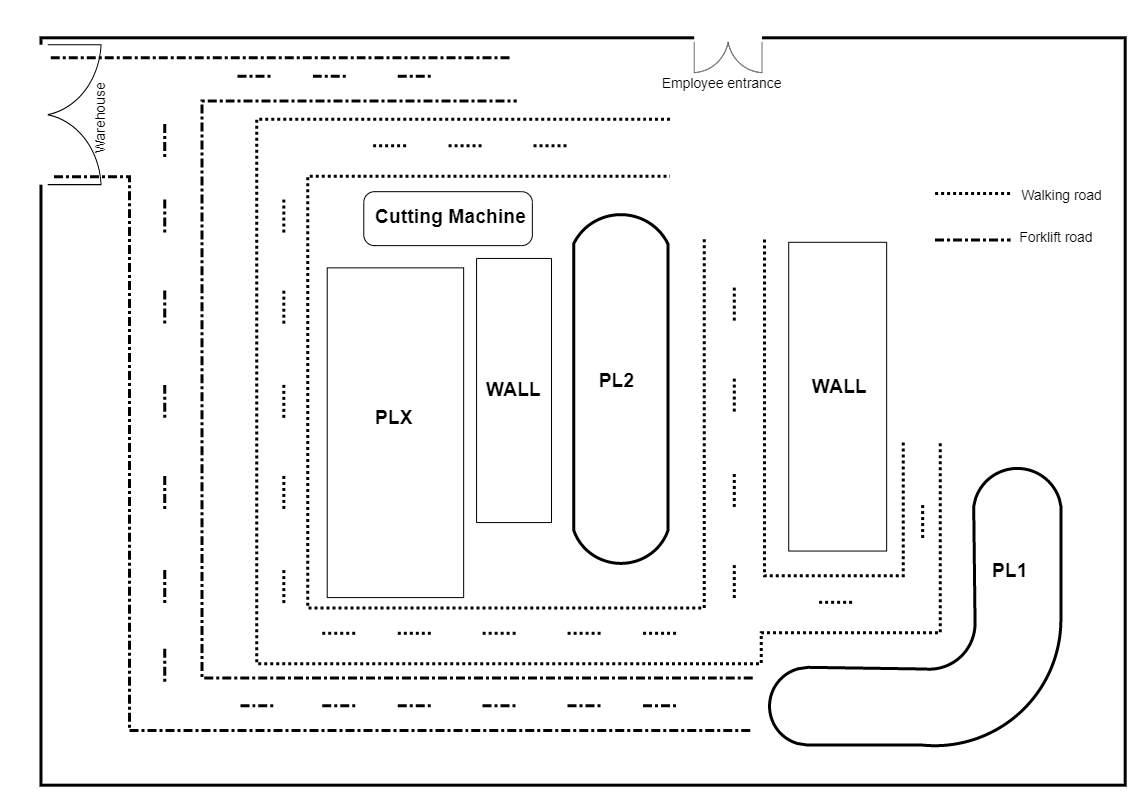}
\end{center}
\caption{Facility layout for felt production}
\label{fig:layout}
\end{figure}

\newpage

\begin{figure}[!h]
\begin{center}
\begin{equation*}
\mathbf{A}=\begin{bmatrix}
1 \ 0 \ 0 \\
1 \ 0 \ 1 \\
1 \ 1 \ 0 \\
1 \ 1 \ 1 \\
\end{bmatrix} \ \ \
\mathbf{P}=\begin{bmatrix}
1 \ 0 \ 0 \\
1 \ 3 \ 0 \\
1 \ 2 \ 0 \\
1 \ 2 \ 3 \\ 
\end{bmatrix}
\end{equation*}
(a) \hspace{10mm} (b) \vspace{-3mm}
\end{center}
\caption{(a) Production route matrix $\mathbf{A}_{im}$, (b) production sequence matrix $\mathbf{P}_i$ of product $i$}
\label{fig:HandGMatrix}
\end{figure}
\par

Figure 2a shows product-machine assignment as $\mathbf{A}$ matrix. For example, chemical cylinders ($i=3$) are processed in PL1 ($m=1$) and PL2 ($m=2$), In the $\mathbf{A}$ matrix this is shown as 1 in the corresponding $(i,m)$ cells, i.e.,   $\mathbf{A}_{31}$ $\mathbf{A}_{32}$ 1 and $\mathbf{A}_{33}$ is 0, since the chemical cylinder ($i=3$) does not go through the cutting operation. Whereas in Figure 2b, shows the precedence relationship $\mathbf{P}$ of the products in particular, $\mathbf{P}_i$ represents the sequence of the machines that product $i$ visits respectively. 

For computational simplicity, the machine speeds are converted to the unit processing time of a machine, which simply shows the time it takes to produces one unit of a product in the unit of minute per unit. The unit processing time on machine $m$ is limited by the lower bound $v^{min}_m$ and the upper bound $v^{max}_m$. The unit processing time and the machine speed are inversely proportional. That is as the unit processing time decreases the machine speed increases. High machine speeds consume more energy and the energy cost of the system increase in parallel. Hence, we assume that all machines are operating at $v^{min}_m$ and aim to optimize the machine speeds to decrease the energy consumption. Cost of energy consumed by the machine $m$ in one minute is represented as $r_m$. Total energy consumption of the machine $m$ in a production cycle is the multiplication of $r_m$ with the unit processing time. Daily production capacity of the machines is the same for every machine and expressed as $m_t$. 


In every period $t$, there is a demand for each product type ($d_{it}$) that must be satisfied throughout the planning horizon $T$. At the end of the periods, it is possible to keep products as end item in the warehouse with the cost of $h_i$ for satisfying the demand in the following periods. Alternatively, it is practical to keep work-in-process (WIP) inventory at the warehouse with the cost of ($w_i$) between the periods. In the case of inventory accumulation, forklifts transport the products to the warehouse from the factory level, generating the inventory transportation (IT) cost ($\bar{t}_i$).

\begin{onehalfspace}
\footnotesize
\begin{center}
\captionof{table}{List of the parameters}
    \label{tab:lop}
\begin{tabular}{l p{0.36\textwidth} l l} 
\hline
\multicolumn{4}{c}{\textit{Parameters}} \\
\cline{1-4}
$\mathbf{I}$ & set of products & $r_m$ & operating cost of machine $m$  \\
$I$ & number of products & $\bar{t}_i$ & IT cost of product $i$  \\
$\mathbf{T}$ & set of periods & $c_{m}$ & VAO made by machine $m$ \\
$T$ & number of periods (planning horizon) & $h_i$  & end item IH cost of product $i$   \\ 
$\mathbf{M}$ &  set of machines & $w_i$  & WIP holding cost of product $i$  \\
$M$ & number of machines & $d_{it}$ & demand of product $i$ in period $t$  \\
$\mathbf{A}_{im}$ & production route matrix & $m_t$ & plant capacity of the factory in period $t$ \\
$\mathbf{P}_i$& production sequence matrix & $u^{max}$ & WIP inventory capacity \\ 
$v^{min}_m , v^{max}_m$&lower and upper bounds on the & $s^{max}$ & end item inventory capacity \\ 
&unit processing time of machine $m$ & $L$ &  a unit of product in meters\\
\hline
\end{tabular}
\end{center}
\end{onehalfspace}
\vspace{4mm}
For formulation simplicity, we assume that production cost of a product is distributed into the machines. In other words, production cost of the products consist of all the value added operations (VAO) that made by that machine. For example, in PL2 there is a chemicalization and drying processes to compact felt. With these processes, basic compact felt is turned to insulated felt. These operations as a whole adds value to the product. Raw materials that is consumed and the machining parts that is used in these processes are shown as $c_m$, VAOs that happened on the product $i$ by machine $m$. By this adjustment, machines and their operations generate cost, not the product type as a whole.

The proposed production plan aims to minimize the total cost by decreasing the inventory levels, deciding what to produce when to produce all by dynamically changing the machine speeds (unit processing times of the machines) at the start of the each period $t$, throughout the planning horizon $T$. We aim to achieve just-in-time (JIT) production at the end of the production. 


This paper is organized at follows: Section 3 explains the followed solution method in detail. Section 3.1 explains the LSMS model and Section 3.2 describes the developed \emph{Two-Phase} Heuristic. Section 4 shows computational results of the \emph{Two-Phase} Heuristic performance according to the generated scenarios. The behaviour of the system according to the unit processing time and demand is analyzed explained at Section 4.1, the total production and inventory explained in Section 4.2. Section 5 concludes the paper together with some comments on the future work.

\section{Solution Method}

We first propose a NLP formulation for the LSMS problem in Section 3.1. The problem complexity of the LSMS problem is discussed in the same section. Then to produce near-optimal solution in polynomial time, our \emph{Two-Phase} Heuristic is introduced in Section 3.2.


The model we proposed is NLP which will be discussed in Section 3.1. With the aim of solving the model in a polynomial time, we created a heuristic named as \emph{Two-Phase} which will be explained in Section 3.2.



\subsection{Mathematical Formulation}

We proposed an NLP model with the objective of minimizing the total cost while taking into account unit processing time, plant capacity and inventory constraints of the felt production system. In this production plan we allowed to change the machine speeds, in other words, unit processing times of the machines by overloading the machines in a period within the acceptable boundaries. Decreasing the unit processing time of the machine (increasing the machine speed) from its original time generates a penalty which corresponds to the energy consumption of the machine. By this idea, we proposed our mathematical model with consideration of changing the unit processing times of the machines at the start of each period. 

\begin{onehalfspace}
\footnotesize
\begin{center}
\captionof{table}{List of decision variables}
    \label{tab:lop}
\begin{tabular}{l p{0.36\textwidth} l l} 
\cline{1-4}
\multicolumn{4}{c}{\textit{Decision Variables}} \\
\cline{1-4}
$v_{mt}$ & unit processing time on machine $m$ & $y_{imt}$ & amount of production of product $i$  \\
&  in period $t$ & & on machine $m$ in period $t$\\
$s_{it}$ & inventory level of product $i$ & $u_{it}$ & WIP inventory level of product $i$ \\
&  at the end of period $t$ & & at the end of period $t$\\
\hline
\end{tabular}
\end{center}
\end{onehalfspace}
\vspace{4mm}
Objective function (\ref{Objective}) consists of summing all cost which are production, inventory holding and transportation cost then subtracting machine speed penalty within a period until the end of the planning horizon. Normally, we want to add the machine speed penalty to the objective. But we have $v_{mt}$ unit processing time as decision variable which is inversely proportional to the machine speed. Hence, in the objective machine speed penalty appears with a minus sign. With this modification of the penalty, we ensure that the unit processing times of the machines change according to the demand requirements of the periods. 

\textbf{Lot Sizing and Machine Speed (LSMS) Model:}
\begin{align}
 \min z = & \sum_{t \in T}c_my_{imt} + \sum_{i \in I} \left\{ s_{it}(h_i + \bar{t}_i) + u_{it}(w_i + \bar{t}_i) \right\}  - \sum_{m \in M} r_mv_{mt} \label{Objective} \\
 &\sum_{i \in I} y_{imt}v_{mt} \leq m_t \hspace{32mm}  \forall m \in M, \quad \forall t \in T \label{capacity}\\
 & y_{imt} \leq \left(\sum_{t \in T} d_{it}\right)\mathbf{A}_{im} \hspace{28mm}   \forall i \in I \quad \forall t \in T \label{totaldem}\\ 
 & \sum_{t \in T}y_{i1t} = \sum_{t \in T}y_{i2t} \hspace{36mm}   \text{ for } \hspace{1.5mm}  i \in  \{3,4\} \label{flow}\\
 & s_{it-1} + y_{im^*t} = d_{it} + s_{it} \hspace{27mm}  \forall i \in I, \hspace{1.5mm} \forall t \in T \label{endbalance} \\
 & u_{2t-1} + y_{21t} = y_{23t} + u_{2t}  \hspace{33mm} \forall t \in T \label{wipbalance1} \\
 &  u_{4t-1} + y_{42t} = y_{43t} + u_{4t}  \hspace{33mm}  \forall t \in T \label{wipbalance2} \\
  & v^{min}_m \leq v_{mt} \leq  v^{max}_m \hspace{32mm}  \forall m \in M, \quad \forall t \in T \label{bounds}\\
  &\sum_{i \in I}s_{it} \leq s^{max} \hspace{50mm}  \forall t \in T \label{maxend}\\
  &\sum_{i \in I}u_{it} \leq u^{max} \hspace{50mm}  \forall t \in T \label{maxwip} \\
  & u_{it} = 0  \hspace{51mm}  \forall t \in T, \text{ for } \hspace{1.5mm} i \in  \{1,3\} \label{nowip}
\end{align}

In constraints (\ref{capacity}) the total production time in a given machine must not exceed the available time $m_t$. Constraint (\ref{capacity}) are nonlinear due to the multiplication of the decision variables $y_{imt}$ and $v_{mt}$. This makes the LSMS formulation a nonlinear model.

In the product-machine $\mathbf{A}$ matrix, cells can only take binary values indicating the presence of the production. The value of $\mathbf{A}_{32}$ is one, meaning that there is a production of the product $i=3$ in the machine $m=2$. The value of $\mathbf{A}_{33}$ is zero, showing that there is no production of the product $i=3$ in the machine $m=3$. By multiplying the total demand with $\mathbf{A}$ matrix in constraint (\ref{totaldem}), we ensure that only the feasible product types at given machines can be produced as well as allowing machines to produce more products than that periods demand to keep inventory if necessary.



\begin{lemma} The LSMS model is a non-convex nonlinear model. 

\textbf{Proof.} In order to have a convex mathematical model, the objective function and the equality constraints should be linear and the less-than-or-equal-to constraints should be convex. The constraints (2) include decision variables $y_{imt}$ and $v{mt}$ multiplied. That is, these constraints are not convex which results that the LSMS model to be non-convex. $\square$
\end{lemma}

One of the main assumptions of the production system is that there will be no WIP products from $m=1$ to $m=2$. Idea behind this assumption is that products that unloaded from $m=1$ can be loaded on $m=2$ via crane system which is embedded between that two production lines. Constraint (\ref{flow}) assure that, total chemical type $(i = 3, 4)$ production in $m=1$ (PL1) is equal to total chemical type production in $m=2$ (PL2). Constraints (\ref{flow}) show that the chemical products can not start their production on $m=2$ unless they have been produced in $m=1$ first. At the same time, with constraint (\ref{flow}), system does not generate any WIP products between $m=1$ and $m=2$. 


Some periods may be overloaded with demand, which results in exceeding the daily capacity of the production system. To reduce this overload on the production system before these periods it is possible keep the excessive production as end item inventory. Constraints (\ref{endbalance}) are the end item inventory balancing constraints for each product type at the end of a period. All products become end product when they finish their machine routing in their sequence matrix $\mathbf{P}$. To guarantee keeping only end item products as end item inventory, the decision variable $y_{imt}$ is written as $y_{im^*t}$, where $m^*$ is the last machine in the production sequence matrix $\mathbf{P}_i$.

It is possible to keep WIP inventory which only consist of plaque type $(i = 2, 4)$ of products on machine $m=1$ to $m=2$ and $m=3$. Constraints (\ref{wipbalance1}) and (\ref{wipbalance2}) show that the WIP inventory balances at the end of each period. Constraints (\ref{nowip}) ensure that no WIP inventories is generated between the machines $m=1$ and $m=2$.



Each machine in the production system has their unit processing time independent from the product type that they produce, and their unit processing time bounds, which can be seen from Constraints (\ref{bounds}), representing the lower and upper bounds of the unit processing time of the machines.


Constraints (\ref{maxend}) and (\ref{maxwip}) show the total allowed end and WIP item inventories in a period, respectively. Since it is assumed that cylinder type and plaque type of products have the same volume, there is no need to separate product types in the summation. 

\subsection{Two-Phase Heuristic}

As stated earlier, constraints (\ref{capacity}) make the LSMS model a nonlinear problem due to the multiplication of the decision variables, $y_{imt}$ and $v_{mt}$. By using the divide-and-conquer idea we developed the \emph{Two-Phase} heuristic, which divides the nonlinear problem into two linear subproblems (SP) and solves them repeatedly.


In the implementation of the \emph{Two-Phase} heuristic, we divide the LSMS model into two  subproblems namely Subproblem 1 (SP$_1$) and Subproblem 2 (SP$_2$). Each SP has the same objective function, minimizing the cost. In particular SP$_1$ has $y_{imt}, s_{mt}, u_{mt}$ as decision variables and takes the initial unit processing times of the machines  $v_{mt}$ as parameters in the production system. Similarly, SP$_2$ takes $v_{mt}, s_{mt}, u_{mt}$ as decision variables, and keeps the production amount of the machines $y_{imt}$ as parameters. Each model has constraints that are subjected to their decision variables. Mathematical formulation of SP$_1$ can be seen as follows. 

\underline{\textbf{SP$_1$:}}
\begin{align}
 \min z = & \sum_{t \in T}c_my_{imt} + \left(\sum_{i \in I}  s_{it}(h_i + \bar{t}_i) + u_{it}(w_i + \bar{t}_i) \right) - \sum_{m \in M} r_m\hat{v}_{mt} \label{Objectivesp1} \\
 & \sum_{i \in I} y_{imt}\hat{v}_{mt} \leq m_t \hspace{35mm} \forall m \in M, \quad \forall t \in T \label{capacitysp1} \\
 &(\ref{flow}) - (\ref{wipbalance2}), (\ref{maxend}) \text{ and }  (\ref{maxwip}) \nonumber
\end{align}

SP$_1$'s main difference can be seen above. LSMS model is modified according to the $y_{imt}, s_{mt}, u_{mt}$ decision variables. $\hat{v}_{mt}$ is introduced as a parameter which holds the machine unit processing time which is obtained from SP$_2$'s solution or can be given as an initial machine unit processing time in the first execution. With this alteration, now SP$_1$ is a linear model since constraints (\ref{capacitysp1}) are no more nonlinear. 

\underline{\textbf{SP$_2$:}}
\begin{align}
 \min z = &\sum_{t \in T}c_m\hat{y}_{imt} + \left(\sum_{i \in I}  s_{it}(h_i + \bar{t}_i) + u_{it}(w_i + \bar{t}_i) \right) - \sum_{m \in M} r_mv_{mt} \label{Objectivesp2} \\
 & \sum_{i \in I} \hat{y}_{imt}v_{mt} \leq m_t \hspace{35mm} \forall m \in M, \quad \forall t \in T \label{capacitysp2}\\
  & s_{it-1} + \hat{y}_{im^*t} = d_{it} + s_{it} \hspace{29mm}  \forall i \in I, \quad \forall t \in T \label{endbalancesp2} \\
 & u_{2t-1} + \hat{y}_{21t} = \hat{y}_{23t} + u_{2t}  \hspace{37mm} \forall t \in T \label{wipbalance1sp2} \\
  & u_{4t-1} + \hat{y}_{42t} = \hat{y}_{43t} + u_{4t}  \hspace{37mm}  \forall t \in T \label{wipbalance2sp2} \\
  & (\ref{bounds}) - (\ref{maxwip}) \nonumber
\end{align}

The SP$_2$, LSMS model is modified according to the $v_{mt}, s_{mt}, u_{mt}$ decision variables. $\hat{y}_{imt}$ is introduced as a parameter which holds the production amounts which is obtained from the solution of SP$_1$. As in SP$_1$, SP$_2$ is a linear model since constraints (\ref{capacitysp2}) are linear.



\begin{figure}[!h]
\begin{center}
\includegraphics[width=1\columnwidth]{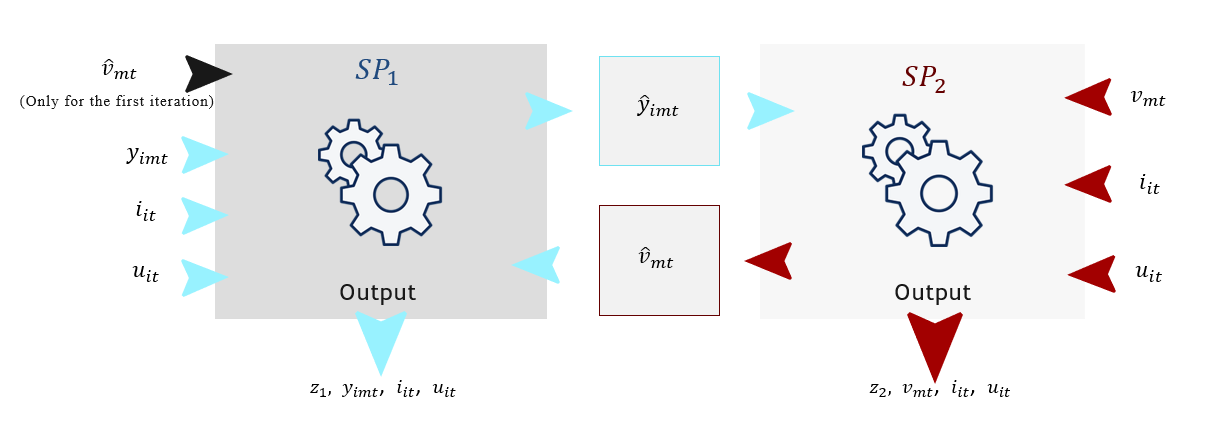}
\end{center}
\caption{Visualization of the \emph{Two-Phase} Heuristic}
\label{fig:twophase}
\end{figure}

To solve the LSMS model, subproblems need to communicate with each other. To solve SP$_1$, we give initial unit processing times of the machines ($v_{mt}$) as parameters. With this initial processing times, SP$_1$ is executed and feasible values for the $y_{imt}, s_{mt}, u_{mt}$ decision variables are found. As explained earlier, from these values $y_{imt}$ is given to the SP$_2$ as a parameter. Then, SP$_2$ is executed and feasible values for the $v_{mt}, s_{mt}, u_{mt}$ decision variables are obtained. One cycle of \emph{Two-Phase} consists of solving SP$_1$ \& SP$_2$ once. At the end of the each cycle, the objective function values of the SP$_1$ \& SP$_2$ compared with each other. If the objective function values are the same, the algorithm compares $s_{mt}, u_{mt}, y_{imt}, v_{mt}$ decision variable values from current and previous cycles. This comparison will continue as long as there is no difference in the decision variables and the objective functions. Figure \ref{fig:twophase} visualizes the cycle of the \emph{Two-Phase} heuristic. Pseudocode of the \emph{Two-Phase} heuristic is provided below (see Algorithm 1).

\vspace{-2mm}
\begin{center}
\onehalfspacing
\footnotesize
$
\begin{tabular}{ll}
\textbf{Algorithm 1:} ($\emph{Two-Phase}$) \\ 
\hline
\textbf{Input:} An instance of $LSMS$ model\\ 
\hline
0. $loop \leftarrow true$ \\ 
1. $\hat{v}_{mt}$ $\leftarrow$ $v^{min}$ for all $m, t$ \\
2. \textbf{While}  $loop \leftarrow true$ \\ 
3. \hspace{20pt} Initialize \emph{condition}.\\ 
4. \hspace{20pt} Save previous $y_{imt}, v_{mt}$ . \\
5. \hspace{20pt} Run SP$_1$ with $\hat{v}_{mt}$, for $\mathbf{I,M, T}$ sets, find current $y_{imt}, s_{it}, u_{it}$  \\
6. \hspace{20pt} Save $\hat{y}_{imt}$ as ${y}_{imt}$  \\
7. \hspace{20pt} Run SP$_2$ with $\hat{y}_{imt}$, for $\mathbf{I, M, T}$ sets, find current $ v_{mt}, s_{it}, u_{it} $  \\
8. \hspace{20pt} $\hat{v}_{mt}$ $\leftarrow$ ${v}_{mt}$  \\
9. \hspace{20pt} \textbf{If} there is change in $y_{imt}, v_{mt}, s_{it}, u_{it}$ \textbf{Then} \\
10. \hspace{30pt} $condition \leftarrow true$\\
11. \hspace{20pt} \textbf{Else} $condition \leftarrow false$ \\
12.\hspace{20pt} \textbf{End If}\\
13.\hspace{20pt} \textbf{If} $condition \leftarrow false$ \textbf{Then} Go to Step 2.\\ 
14.\hspace{20pt} \textbf{Else If} $condition \leftarrow true$ \textbf{Then} $loop \leftarrow false$, Go to Step 2\\
15.\hspace{20pt} \textbf{End If}\\
16. \textbf{End While}\\
\hline
\textbf{Output:} Near-optimal solution of the LSMS model\\$y_{imt}, v_{mt}, s_{it}, u_{it}$ values with objective value $Z$\\
\hline
\end{tabular}
$
\end{center}

If $T_{max}$ is the maximum iteration limit in the \emph{Two-Phase} heuristic, then its time complexity can be given as $\mathcal{O}$((SP$_1$ + SP$_2$)$T_{max}$) where  $\mathcal{O}$(SP$_1$) and  $\mathcal{O}$(SP$_2$) are the time complexity of SP$_1$ and SP$_2$, respectively. Note that $\mathcal{O}$(SP$_1$) and  $\mathcal{O}$(SP$_2$) are polynomial time complexity since SP$_1$ and SP$_2$ are linear programming problems

\section{Computational Results}

Computational parameters gathered by examining the production system as well as making a market research for the production costs. The system is analyzed for three different planning horizon levels, i.e. $T$ = 10, 20 and 30 periods. For each $T$, 10 random demand instances are generated. For better understanding the effects of the parameters on the objective function and the production system, two key parameters $m_t$ and $(s^{max}_i, u^{max}_i)$, which effect the decision variables the most, are chosen and created three different scenarios for each parameter. These scenarios can be explained as Low, Medium and High level instances, which are in total of 270 instances. List of computational parameters can be seen in Table 2. 

\begin{onehalfspace}
\begin{center}
\captionof{table}{List of the computational parameters}
    \label{tab:locp}
\begin{tabular}{l l}  
\hline
$c_m$ & [2400, 5400, 1000] \\
$\bar{t}_i$ &  [100, 120, 120, 140] \\
$h_i$ & [300, 150, 300, 150] \\
$w_i$ & [0, 50, 0, 50] \\
$r_m$ & [1.16, 3.09, 0] \\
$T$ & 10, 20, 30 \\
$L$ & 400 meter\\
$m_t$ & 630, 720, 810 minutes\\
$(s^{max}_i, u^{max}_i)$ & (6,3), (12,6), (18, 9)\\
$D_{T}$ & (80,120), (180,250), (290,350)\\
$(I, M)$  & (4,3) \\
\hline
\end{tabular}
\end{center}
\end{onehalfspace}

The felt company provided us a demand ratio for each product type and gave us a minimum and maximum amount of demand that can be requested in a given planning horizon $T$. For planning horizons $T=10$, $T=20$ and $T=30$ total demand bounds are (80, 120), (180, 250) and (290, 350) respectively. From these total demands:
\begin{itemize}
    \item 16\% of the total demand corresponds to demand of the product $i=1$.
    \item 4\% of the total demand corresponds to demand of the product $i=2$.
    \item 64\% of the total demand corresponds to demand of the product $i=3$.
    \item 16\% of the total demand corresponds to demand of the product $i=4$.
\end{itemize}

With the data provided, demands are generated and distributed on the periods within the given planning horizon randomly using uniform distribution. VAOs of the machines, $c_m$, represents an array of single instance. In this instance, $c_1$ is 2400, $c_2$ is 5400 and $c_3$ is 1000. Similar to $c_m$, $\bar{t}_i$, $h_i$, $w_i$ and $r_m$ represents an array of single instances which changes with the corresponding index values. These computational parameter values are generated by analysis of the production system.

A unit of product corresponds to a 400 meters of felt. It is assumed that all product types (plaque and cylinder) takes the same amount of space in the inventory, at the same time, holding cost of cylinder type products is higher since it requires more manpower and better allocation in the warehouse. In Table 3, lower and upper bounds for unit processing time of the machines can be seen.

\begin{onehalfspace}
\begin{center}
\captionof{table}{Unit Processing Time Bounds of Machines}
    \label{tab:locp}
\begin{tabular}{l l}  
\hline
$(v^{min}_1, v^{max}_1)$ & (50, 80) min/unit \\ 
$(v^{min}_2, v^{max}_2)$ & (22.2, 26.6) min/unit \\
$(v^{min}_3, v^{max}_3)$ & (80, 80) min/unit \\
\hline
\end{tabular}
\end{center}
\end{onehalfspace}

The computations are conducted on a computer with 2.80GHZ Intel Core i7-7700HQ processor working on Windows 10 with 16GB RAM. All codes are written with Java language, on Eclipse IDE. For generating the optimal solution of the subproblems we utilized IBM CPLEX 12.10. From the outputs only average values reported. Since all demands are satisfied throughout the planning horizon, $\bar{y}_{imt}$ would be equal to $\bar{d}_{it}$ and would be the same from instance to instance, therefore no $y_{imt}$ value is reported. We reported only $\bar{v}_{1t}$, the average unit processing time of PL1, in the Tables 4 and 5. Since $m=1$ is the critical production line with producing all the product types. At the same time, in all instance outputs, ${v}_{2t}$ value is constant and equal to 26.6 min/unit and ${v}_{3t}$ is a constant and equal to 80 min/unit.

The reason behind that only $\bar{v}_{1t}$ is reported in the table is that, $m=1$ is the critical production line with producing all the product types, as well as in 270 instance outputs, ${v}_{2t}$ value is constant and equal to 26.6 min/unit and ${v}_{3t}$ is a constant and equal to 80 min/unit.

For a given $T$ planning period, as the available working hours $m_t$ in a day get higher, the demand can be satisfied with a lower average machine $m = 1$ speed, i.e., $\bar{v}_{1t}$ processing times increase, while accumulating fewer WIP and end-item inventories. 

For all $T$ levels, at  $m_t$(\emph{low}) working hours allowing more WIP and end-item inventories, i.e., $s_{it}$ and $u_{it}$ get higher, decreases the end-item inventory $s_{it}$ amounts and increases WIP inventory $u_{it}$ level. This is since WIP inventory carrying cost is smaller than the one for end-item inventory. 

\begin{onehalfspace}
\begin{center}
\footnotesize
\captionof{table}{Performance of the Two-Phase Heuristic}
    \label{tab:EMDvsBP}
    \scalebox{0.95}{ 
\begin{tabular}{rrcrrrrrr}
    \hline
 &  & & & \multicolumn{4}{c}{Two-Phase} \\ \cline{4-9} 
T & $m_t$ & ($s^{max},u^{max}$)& $\bar{z}$ & CPU (ms)  & \# iter & $\bar{s}$ & \hspace{-3mm} $\bar{u}$  & \hspace{-3mm}$\bar{v}_{1t}$ \\
    \hline
     &  & low & 737,425.85 & 165.84 & 2 & 0.15 & 0.76 &  59.30  \\ 
     & low & med & 737,319.85 & 164.2 & 2 & 0.04 & 0.87 & 59.30   \\
     &  & high & 737,319.85 & 163.22 & 2 & 0.04 & 0.87 & 59.30 \\
     \cline{3-9}
     &  & low & 735,703.11 & 167.58 & 2 & 0.01 & 0.09 & 66.15  \\
   10  & med & med & 735,703.11 & 164.12 & 2 & 0.01  & 0.09  & 66.15   \\
     &  & high & 735,324.26 & 158.92 & 2 & 0 & 0.03 & 71.70  \\
     \cline{3-9}
     &  & low & 735,324.26 & 160.30 & 2 & 0 & 0.03 &  71.70\\
     & high & med & 735,324.26 & 161.80 & 2 & 0 &0.03 & 71.70\\
     &  & high & 735,324.26 & 158.92 & 2 & 0 & 0.03 & 71.70 \\
     \cline{2-9}
     \cline{2-9}
     &  & low & 1.450,956.60 & 185 & 2 & 0.07 & 0.47 &  60.67 \\ 
     & low & med & 1,453,682.87 & 183.2 & 2 & 0.04 & 0.51 & 60.67 \\
     &  & high & 1,456,684.13 & 182.52 & 2 & 0.03 & 0.51 & 60.67\\
     \cline{3-9}
     &  & low & 1,451,398.82 & 183.58 & 2.1 & 0 & 0.05 & 67.28\\
   20  & med & med & 1,451,398.82 & 180.84 & 2 & 0  & 0.05  & 67.28 \\
     &  & high & 1,451,398.82 & 181.88 & 2 & 0 & 0.05 & 67.28 \\
     \cline{3-9}
     &  & low & 1,450,799.37 & 183.5 & 2 & 0 & 0.01 &  72.11 \\
     & high & med & 1,450,799.37 & 181.88 & 2 & 0 &0.01 &  72.11\\
     &  & high & 1,450,799.37 & 180.48 & 2 & 0 & 0.01 & 72.11 \\
    \cline{2-9}
     \cline{2-9}
     &  & low & 2,150,772.64 & 214.2 & 2.3 & 0.17 & 0.50 &  61.21  \\ 
     & low & med & 2,158,481.07 & 200.86 & 2 & 0.04 & 0.55 & 59.95 \\
     &  & high & 2,158,432.18 & 201.78 & 2 & 0.03 & 0.56 & 59.95\\
     \cline{3-9}
     &  & low & 2,154,607.06 & 200.2 & 2 & 0 & 0.07 & 66.01 \\
   30  & med & med & 2,154,607.06 & 202.28 & 2 & 0  & 0.07  & 66.01\\
     &  & high & 2,154,607.06 & 199.56 & 2 & 0 & 0.07 & 66.01\\
     \cline{3-9}
     &  & low & 2,153,550.94 & 200.88 & 2 & 0 & 0 &  71.22 \\
     & high & med & 2,153,550.94 & 198.1 & 2 & 0 &0 & 71.22\\
     &  & high & 2,145,411.34 & 197.62 & 2 & 0 & 0 & 72.79\\
    \hline
\end{tabular}}
\end{center}
\end{onehalfspace}

There is no significant increase in the computation times as the instance gets larger (for the parameters $T, m_t, s^{max}, u^{max}$ from \emph{low} to \emph{high} level). For each instance, all the machines start with their highest production speed, i.e., the smallest $v_{mt}$ value, at the beginning of the algorithm. Under this setting, we observe that the average number of iterations to converge is 2 for almost all instances. 

As $T$ planning horizon gets longer, the total demand for the items increases in the instances as well. Then, we observe an increase in the average objective function value $z$ with the increase in $T$ as expected. 

We observe the best objective function $z$ values, i.e., the smallest production costs, for the  $m_t$(\emph{high}) levels and inv(\emph{high}) bounds. This is since we decrease the WIP and end-item inventories by higher average processing times $\bar{v}_{1t}$ and the operating costs improve with the increased machine speed and the lower inventories in the system. 

The computational experiments do not show any change in the unit processing times of machines $m = 2$ (PL2) with 26.6 min/unit, and $m = 3$ (CM) with 80 min/unit. Hence, in Table 5 we only reported the machine $m = 1$ unit processing times. 

In Table 5, the average outputs of the 270 scenario executions are summarized. As can be seen, as the $m_t$ increases, ($s^{max},u^{max}$) level does not effect the system in terms of objective function value, end item and WIP inventory values and the unit processing time of the PL1. Whereas, in the low levels of $m_t$, ($s^{max},u^{max}$) is the key factor in changes of variables. In all cases, best results can be found in high levels of $m_t$. 

For each $T$, a random demand instance, which is the most overloaded instance with demand, is chosen to visualize its outputs for each scenario. In each graph, columns represents the $m_t$ levels, rows represent the ($s^{max},u^{max}$) levels.

In each $m_t$ level there is a maximum number of products that can be produced in a period which can be found by dividing the $m_t$ level value by the $v^{min}_m$ value. In this case, as we report only $v_{1t}$, this value can be found as $m_t$ divided by $v^{min}_1$. To sum up, for every $m_t$ level, the maximum number of products that can be produced by $m = 1$ is 12.6 unit, 14.4 unit and 16.2 unit, respectively. These bounds can be surpassed only if there is an accumulated WIP inventory from the earlier periods.

There are two types of figures which is used to summarize Table 4. One type is showing the relationship of $\bar{v}_{1t}$ (Unit Processing Time) with $d_t$ (Demand) according to scenarios. In this type red line plot represent the $\bar{v}_{1t}$ value from period to period and its values are shown at the secondary y-axis; bars represents the total demand within a period and its values are shown at the primary y-axis. The other figure type shows the relationship of $y_{imt}$ (Total Production) with $s_{it}, u_{imt}$ (Inventory) According to Scenarios. Each figure type has three rows and three columns. Rows shows the $s^{max},u^{max}$ (inventory) scenarios as \emph{inv(low)}, \emph{inv(med)} and \emph{inv(high)}; columns shows the $m_t$ (daily capacity) scenarios as \emph{mt(low)}, \emph{mt(med)} and \emph{mt(high)}. In total there are nine graphs in a figure which corresponds to intersecting scenarios. Each types of figures are used to summarize each $T$ planning horizon. With total of six figures Table 4 is visualized. From all planning horizons, an instance is chosen with most variability from inventory levels to $v_{mt}$ and $y_{imt}$ levels.

\subsection{The Effects of Unit Processing Time and Demand}

\begin{figure}[!h]
\begin{center}
\includegraphics[width=1\columnwidth]{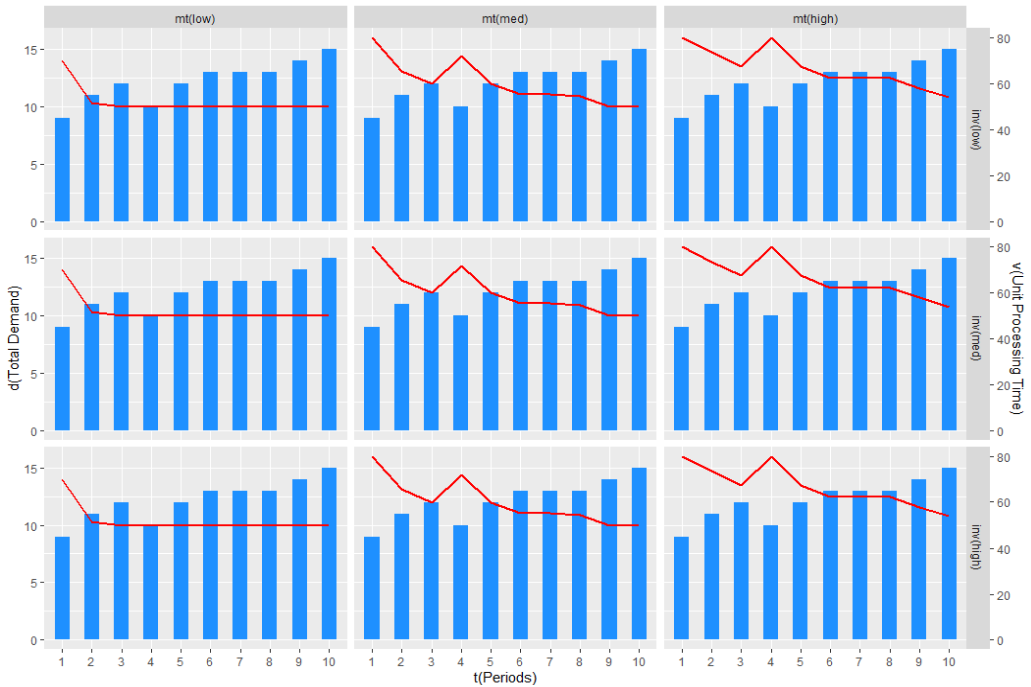}
\end{center}
\caption{Relationship of unit processing time with demand according to the scenarios (T=10)}
\label{fig:T10_speed_demand}
\end{figure}

In the Figure \ref{fig:T10_speed_demand}, $v_{1t}$ (unit processing time of $m=1$) can be seen with the total demand of $d_t$. The graph shows how $v_{1t}$ reacts to a change in demand according to scenarios. As can be seen in the Figure 3, inventory levels do not effect the $v_{1t}$ value. At \emph{mt(low)} level from $t=3$ to $t=10$, $v_{1t}$ value is constant as 50 min/unit, that means that $m=1$ works at its capacity and consuming more energy to satisfy the demand. But at \emph{mt(med)}, $m=1$ can be adjusted with respect to change in demands. As expected, while $d_t$ gets lower, $v_{1t}$ increases ($m=1$ speed decreases) thus consuming less energy; as $d_t$ gets higher, $v_{1t}$ decreases ($m=1$ speed increases) thus consuming more energy. At \emph{mt(high)}, at relatively low $d_t$ periods (relatively low at \emph{mt(high)} $v_{1t}$ value generally close to its upper bound 80 min/unit where energy consumption is low, just like the other $m_t$ levels, as the $d_t$ increases and $v_{1t}$ decreases.

\begin{figure}[!h]
\begin{center}
\includegraphics[width=1\columnwidth]{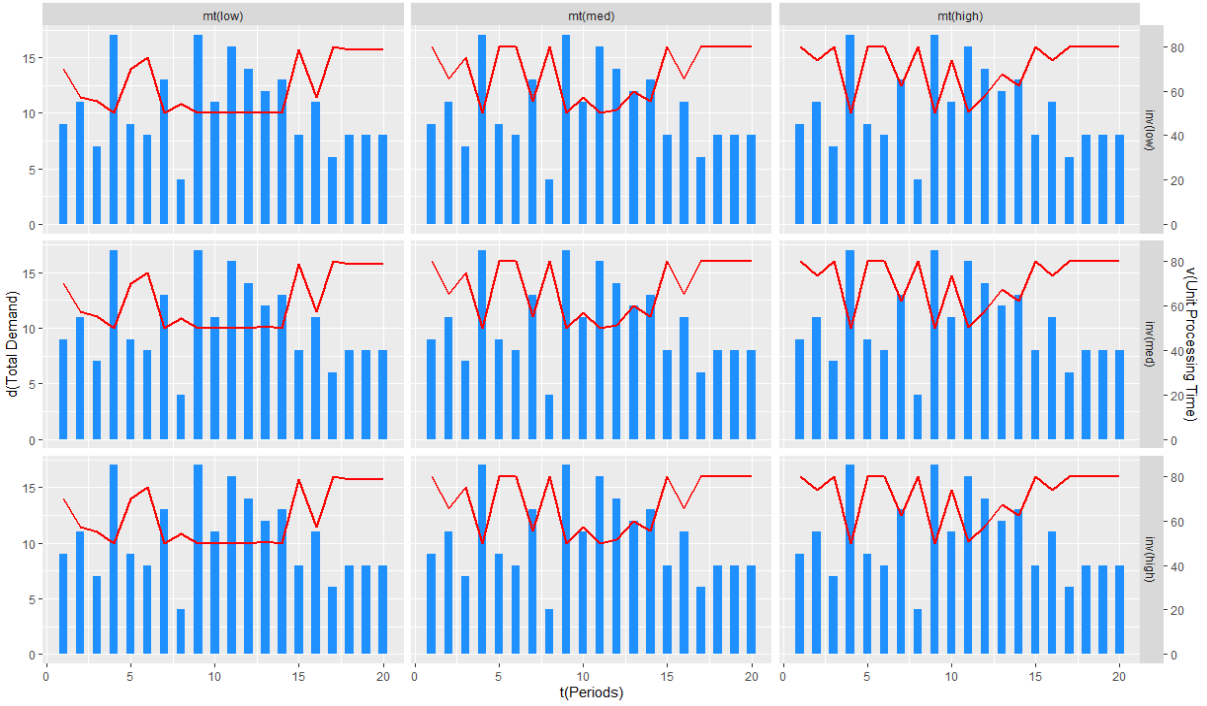}
\end{center}
\caption{Relationship of unit processing time with demand according to the scenarios (T=20)}
\label{fig:T20_speed_demand}
\end{figure}

The same trend continues as the planning horizon becomes $T=20$ . As can be seen in the Figure \ref{fig:T20_speed_demand} just like the $T=10$, only $m_t$ levels affects the $v_{1t}$ values. In all cases, as the $d_t$ increases, $v_{1t}$ tends to decrease, $m=1$ consumes more energy. Although \emph{mt(med)} and \emph{mt(high)} are almost similar in some periods, best scenario, as expected, is \emph{mt(high)}.

\begin{figure}[!h]
\begin{center}
\includegraphics[width=1\columnwidth]{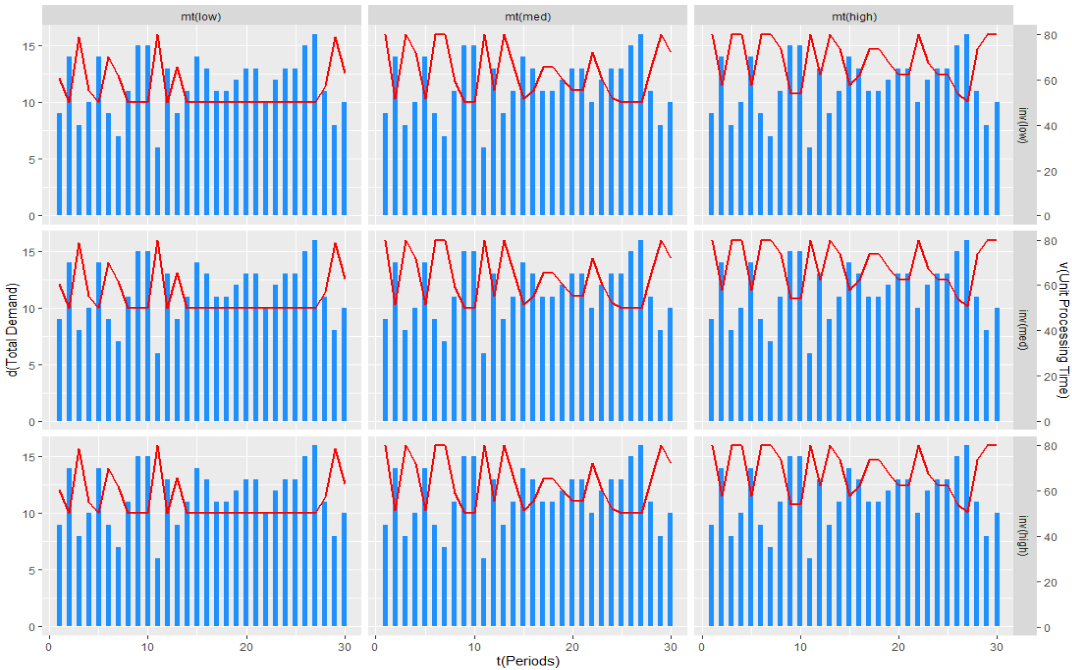}
\end{center}
\caption{Relationship of unit processing time with demand according to the scenarios (T=30)}
\label{fig:T30_speed_demand}
\end{figure}

In Figure \ref{fig:T30_speed_demand}, just like the $T=10$ and $T=20$, as $d_t$ increases, $v_{1t}$ decreases. Similar to lower planning horizons, only $m_t$ level effects $v_{1t}$.

As we look at the general case of relationship of $v_{1t}$ (unit processing time of $m=1$) with the $d_t$ (total demand at period $t$), there is a correlation between them. In all cases regarding the $m_t$ and $s^{max},u^{max}$ levels, as the demand increases unit processing time of $m=1$ tends to decrease, i.e., PL1 starts to work faster and consumes more energy to satisfy the given demand. At all $T$ planning horizons, $m_t$(\emph{high}) generates the best results and $s^{max},u^{max}$ does not effect $v_{1t}$.

\subsection{The Effects of Total Production and Inventory}

\begin{figure}[!h]
\begin{center}
\includegraphics[width=1\columnwidth]{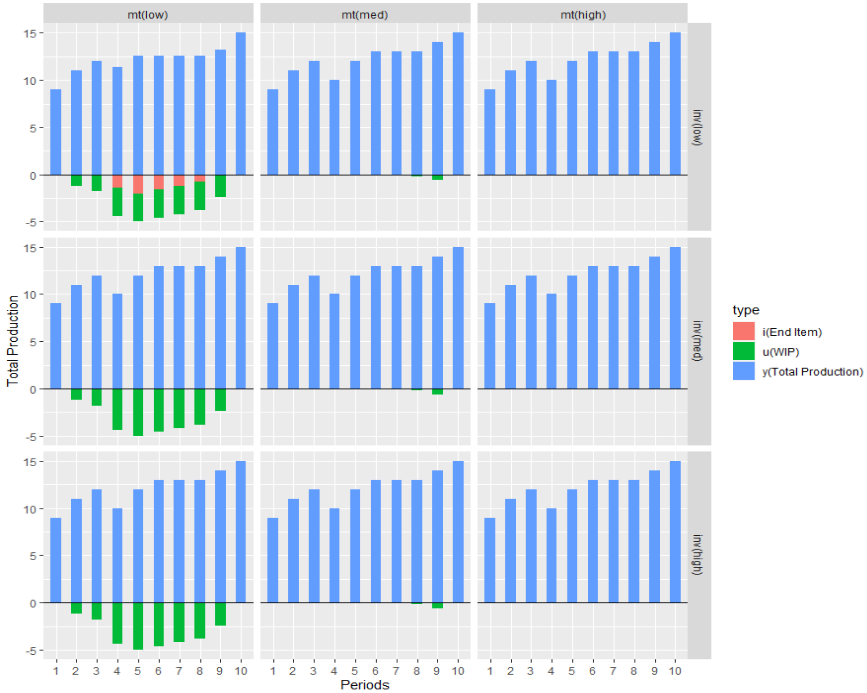}
\end{center}
\caption{Relationship of total production with inventory according to the scenarios (T=10)}
\label{fig:T10_quantity_periods}
\end{figure}

In Figure \ref{fig:T10_quantity_periods}, total production within a period and WIP and end-item inventory levels can be seen. At \emph{mt(low)} and ($s^{max},u^{max}$) levels, accumulation of end item inventory can be seen since there is no more room for the WIP inventory. At \emph{inv(medium)} and \emph{inv(high)} levels of ($s^{max},u^{max}$), there is no end item inventory accumulation. Instead of keeping end item inventory, system decides to keep inventory as a WIP. This shows that keeping WIP inventory is more preferable than keeping end item inventory, also it shows that it is possible to find a feasible solution while keeping end item inventory at low levels of ($s^{max},u^{max}$). As the $m_t$ level increases to \emph{medium}, all inventory levels decrease significantly, thus it becomes possible to satisfy the demand with very-low to zero inventory keeping. At the level of \emph{mt(high)}, it is possible to accomplish just-in-time (JIT) production with zero inventory accumulation on WIP and end-item. 

WIP inventory accumulation does not effect the $y_{imt}$ (total production), it only effects the $y_{imt}$ value only if the system is using the accumulated WIP inventory. For an example, at the start of the accumulation of the WIP inventory, system may be producing 12 units of products, but from these 12 units, 10 of them will become finished goods and 2 of them stored in the inventory as WIP inventory, waiting to be cut. In the graphs, only finished goods are shown as $y_{imt}$. But, as the system starts to use accumulated WIP inventory, WIP products are turned into finished goods thus effects the $y_{imt}$ value. By using WIP inventory, system can exceed its total production capacity in a period. As it comes to the keeping end item inventory, they must be turned into finished goods to become $s_{it}$ (end item inventory). 

\begin{figure}[!h]
\begin{center}
\includegraphics[width=1\columnwidth]{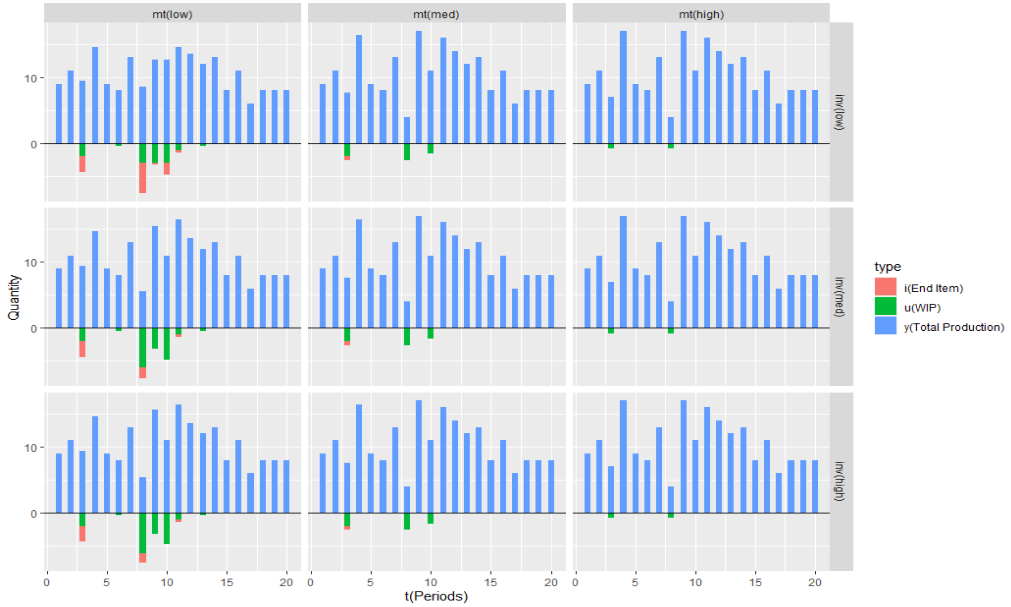}
\end{center}
\caption{Relationship of total production with inventory according to the scenarios (T=20)}
\label{fig:T20_quantity_periods}
\end{figure}

Unlike $T=10$, when planning horizon increases to $T=20$, it may not be possible to satisfy the demand with zero inventory as shown in Figure \ref{fig:T20_quantity_periods}. Although ($s^{max},u^{max}$) inventory levels increase, the system decides to keep some level of end item inventory together with WIP inventory. As a result, even if keeping WIP inventory is more profitable, it can be essential to keep end item as well to satisfy the demand in a period without a backlog. Only at \emph{mt(high)} level there can not be found any end item inventory. In every case, WIP inventory accumulation is unavoidable, but to avoid end item accumulation \emph{mt(high)} must be chosen.

\begin{figure}[!h]
\begin{center}
\includegraphics[width=1\columnwidth]{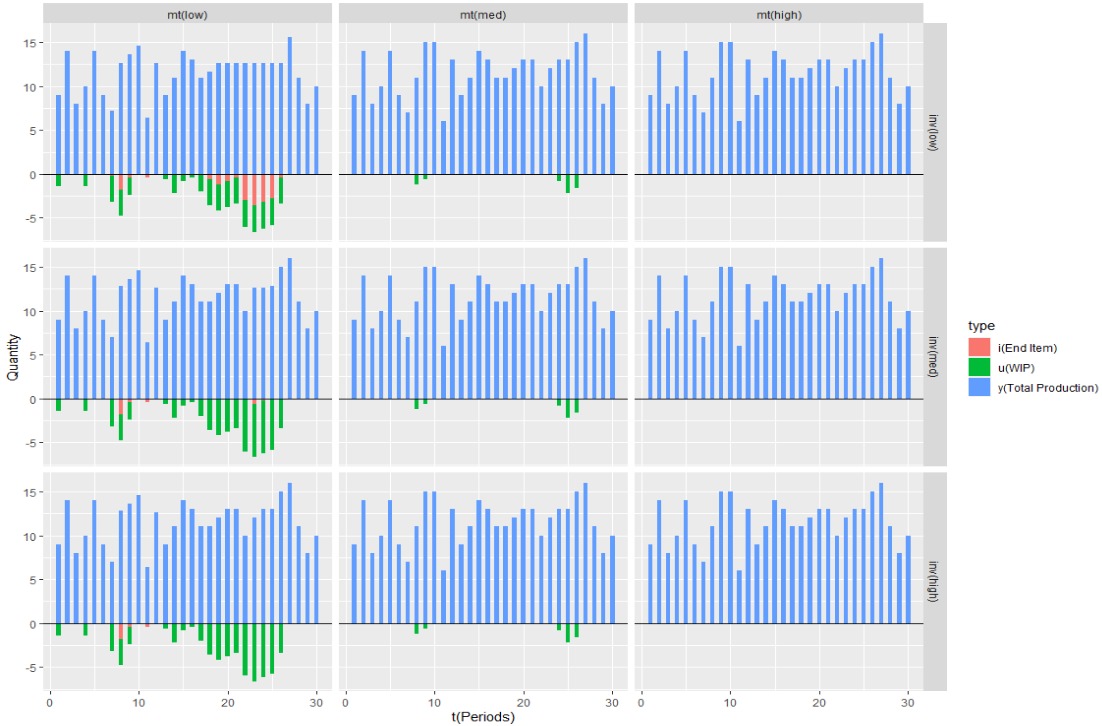}
\end{center}
\caption{Relationship of total production with inventory according to the scenarios (T=30)}
\label{fig:T30_quantity_periods}
\end{figure}


As can be seen in Figure \ref{fig:T30_quantity_periods}, similar to $T=10$ and $T=20$, at \emph{mt(low)} level, ($s^{max},u^{max}$) inventory level play a key role in determining the amount of inventory type kept in a period. As the ($s^{max},u^{max}$) level increases, just like at lower planning horizons, system decreases the amount of accumulated end item inventory and keeps high levels of WIP inventory. At $m_t$(\emph{med}) system only keeps low amounts of WIP inventory and no end item inventory. Similar to $T=10$, system can start to satisfy the demand with no inventory, thus making a JIT production the level is \emph{mt(high)}.

As we look at the general case of relationship of $y_{imt}$ (total production) with $s_{it}$ and $u_{it}$ (end-item \& WIP), we can conclude that only at \emph{mt(low)} level ($s^{max},u^{max}$) affects the system. In all cases, system prefers to keep WIP inventory rather than end item inventory although, in some periods, it becomes necessary to keep end item inventory as well to satisfy the demand throughout the planning horizon. Main parameter that affects the system is $m_t$ since as the $m_t$ level increases, amount of kept inventory significantly decreases. It can be seen that, in some instances, at $m_t$(\emph{high}), it might be possible to do JIT production.


\subsection{The Effects of Daily Capacity and Planning Horizon}

\begin{figure}[!h]
\begin{center}
\includegraphics[width=1\columnwidth]{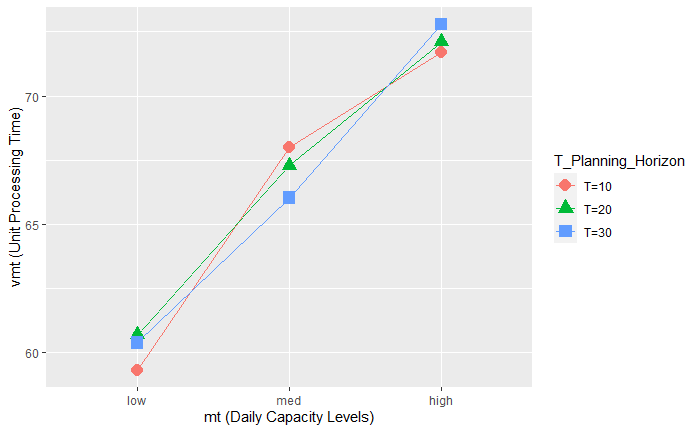}
\end{center}
\caption{Average Processing Times of $m=1$ With Respect to $m_t$ Levels }
\label{fig:averagespeed}
\end{figure}

At Figure \ref{fig:averagespeed}, the trend of average unit processing time of $m =1$ ,i.e., $\bar{v}_{1t}$ can be seen. In all planning horizons, $\bar{v}_{1t}$ tends to increase from level to level as expected. At every $m_t$ scenario, best planning horizon where the average energy consumption is the lowest, i.e., the highest $\bar{v}_{1t}$ changes. At \emph{mt(low)}, 
\emph{mt(med)} and \emph{mt(high)} levels the smallest average energy consumption ($\bar{v}_{1t}$) can be found in $T=20$, $T=10$ and $T=30$ planning horizon levels respectively.


As can be seen, although lowest average energy consumption happens at \emph{mt(high)}, this figure gives an insight about which $T$ planning horizon to choose at given $m_t$ level. 

\section{Conclusions}

In this study we focused on the machine speed and lot sizing problem in a felt production system. We proposed a nonlinear programming (NLP) mathematical model named as lot sizing and machine speed (LSMS) model with the purpose of making a production plan, by optimizing the production amounts, decreasing the inventory levels and by changing the machine speeds which effect the unit processing times of the machines. By changing the machine speeds, we have been able to make a lot sizing, thus creating a production plan which can result in a production system with zero inventory accumulation. To overcome the nonlinearity in the LSMS, we developed a heuristic called \emph{Two-Phase}. By implementing the developed \emph{Two-Phase} heuristic, we solved the NLP model in a polynomial time and obtain a near-optimal solution for the non-convex LSMS. After achieving a feasible solution for the LSMS, we generated scenarios according to the planning horizon, the inventory levels of the inventory types and the total production capacity of the system in a period. We executed the LSMS model in accordance with the generated scenarios. Average outputs and their behaviours of the LSMS, with respect to scenarios, are visualized. Observations showed that, in some scenarios, it is possible to satisfy the demand with no inventory, thus just-in-time production is achievable.

In this work, we generate a \emph{Two-Phase} heuristic for sequentially flowing felt semi-products in a simple three machine flowshop. As a future work, one can consider more complicated flow shop in the production environment with multiple product types.



\newpage
\printbibliography
\end{document}